\documentclass[letterpaper,journal]{IEEEtran}

\ifdefined\XeTeXversion
  \usepackage{fontspec}
\fi

\usepackage{amsmath,amsfonts,amssymb}
\usepackage{array}
\usepackage{algorithm}
\usepackage{algorithmic}
\usepackage{booktabs}
\usepackage{cite}
\usepackage{graphicx}
\graphicspath{{figures/}}
\usepackage{textcomp}
\usepackage{url}
\usepackage{stfloats}
\usepackage{needspace}


\begin{document}

\title{LLM-DR: An On-Demand Delivery Simulation Framework for Studying Heterogeneous Rider Behavior and Collective Mobility with Persona-Grounded Large Language Model Agents}

\author{Chengbo~Zhang, Zekai~Jin, Kaihan~Zhang, Zihan~Yang, Qinyu~Cui, Zuopeng~Xiao, and Jiangbo~Yu%
\thanks{\textit{Corresponding author: Jiangbo Yu.} This work was supported by the Natural Sciences and Engineering Research Council of Canada (NSERC) through a Discovery Grant, the Canada Foundation for Innovation (CFI) through the John R. Evans Leaders Fund (JELF), and the Government of Canada's New Frontiers in Research Fund (NFRF), Exploration stream, an initiative of the Canada Research Coordinating Committee (CRCC).}%
\thanks{Chengbo Zhang, Zekai Jin, and Jiangbo Yu are with the Department of Civil Engineering, McGill University, Montreal, QC, Canada (e-mails: chengbo.zhang@mail.mcgill.ca; zekai.jin@mail.mcgill.ca; jiangbo.yu@mcgill.ca; Chengbo Zhang's ORCID: 0009-0003-7872-4519).}%
\thanks{Kaihan Zhang is with the Cho Chun Shik Graduate School of Mobility, Korea Advanced Institute of Science and Technology (KAIST), Daejeon 34051, South Korea, and the Senseable City Laboratory, Department of Urban Studies and Planning, Massachusetts Institute of Technology, Cambridge, MA, USA (e-mail: kaihn@kaist.ac.kr; ORCID: 0009-0007-1502-571X).}%
\thanks{Zihan Yang and Qinyu Cui are with South China University of Technology, Guangzhou, China, and Ghent University, Ghent, Belgium (e-mails: ctzhyang@mail.scut.edu.cn; ct\_cuiqinyu@mail.scut.edu.cn).}%
\thanks{Zuopeng Xiao is with the Institute of Future Human Habitats, Shenzhen International Graduate School, Tsinghua University, Shenzhen, China (e-mail: tacxzp@foxmail.com).}}

\markboth{IEEE Transactions on Computational Social Systems}%
{Zhang et al.: LLM-DR for Heterogeneous Rider Behavior and Collective Delivery Mobility}

\maketitle
\raggedbottom

\begin{abstract}
On-demand delivery platforms are computational social systems in which dispatch rules and repeated rider actions jointly shape service and urban mobility. Riders choose whether to reject offers, accept one order, or bundle several orders; assignments then change their availability and other riders' opportunities. Modeling this feedback requires empirical rider heterogeneity, shared-order interaction, and road-network execution to be represented together. Delivery optimization and large language model (LLM) mobility generation provide useful foundations, but leave this combination insufficiently examined. We develop Large Language Models as Delivery Riders (LLM-DR), an agentic simulation framework for benchmarking rider policies under a common order stream and dispatch protocol. Persona-grounded LLM agents provide a reference implementation, using empirical profiles, local observations, short-term memory, and route-tool estimates to propose order actions. The environment resolves assignment conflicts, constructs pickup and drop-off sequences, and executes tool-supported routes. Analysis of 9,025 Beijing orders and a sample of 2,508 routed delivery waves identifies four rider personas. A controlled single-day reconstruction places 60 agents, equally distributed across personas, in an observed stream of 8,656 orders. Persona groups complete 16--30 orders per rider, with average bundle sizes of 1.116--2.212. Distributional comparisons retain several empirical persona orderings in order volume, bundling, and routed-wave distance. Fleet-level outcomes show displaced demand, service-occupancy, and completion peaks, alongside a shared central service core and differentiated spatial reach. These results connect data-derived rider strategies to executable action chains and collective mobility patterns within a shared computational social system.
\end{abstract}


\Needspace{4\baselineskip}
\section{Introduction}
\IEEEPARstart{O}{n-demand} delivery platforms coordinate restaurants, customers, and riders through digital dispatch and physical service, forming a computational social system \cite{lazerComputationalSocialScience2020,hofmanIntegratingExplanationPrediction2021}. At each decision, an idle rider chooses whether to reject available offers, accept one order, or bundle several orders. The platform resolves competing acceptances, and assigned orders leave the shared pool. Executing pickup and drop-off actions changes the rider's location and availability, affecting subsequent offers and the opportunities available to other riders. Repeated decisions connect food-service organization \cite{meemkenResearchPolicyFoodDelivery2022,zhaoOnlineOfflineFoodDelivery2021,dengLLM4GKID2026} to urban mobility and spatially evolving delivery networks \cite{zhangUncoveringCommunityStructure2026}. This feedback between human strategies, digital coordination, and physical outcomes provides a setting for empirical-computational studies of cyber-physical-social systems and parallel societies \cite{zhangCyberPhysicalSocialSystems2018,wangParallelSocieties2020}.

Three features make this feedback difficult to represent faithfully in simulation. Riders differ in working times, stacking tendencies, distance tolerance, and spatial reach, consistent with individual variation in human mobility \cite{gonzalezUnderstandingIndividualHuman2008,songLimitsPredictabilityHuman2010}. They also interact through assignments that change the shared order pool and their later exposure to available tasks. Selected orders must become pickup and drop-off actions whose duration and geography depend on the road network and service requirements. Connecting these aspects lets computational experiments examine heterogeneous rider responses and collective mobility within the same system \cite{wangSocialTransportation2019}.

Delivery simulation and optimization have developed approaches to fleet sizing, scheduling, and order assignment \cite{sinhaSimulationbasedStudyDetermine2021,wangOnlineDeepReinforcement2023}. Rider-centered recommendation and crowdsourced order allocation also incorporate individual preferences into platform coordination \cite{wangOnlineDeepReinforcement2023,wangRecommendingGrabbing2023}. These approaches represent rider responses through specified rules, utilities, or learned policies evaluated for their allocation objectives. For behavioral reconstruction, an additional question is whether these representations preserve empirical differences in work timing, bundling, and spatial reach across repeated decisions. The interaction between demand, restaurant geography, and rider types further motivates evaluating behavior within its operating environment \cite{liRealtimeDemandsRestaurant2025}.

LLM mobility models use personal profiles and activity context to generate individual travel diaries or trajectories \cite{wangLargeLanguageModels2024,juTrajLLMModularLLMEnhanced2025,liBeMoreReal2024,shaoChainofPlannedBehaviourWorkflowElicits2024}. Generative-agent research also shows how memory and planning support behavior in interacting populations \cite{parkGenerativeAgentsInteractive2023}. LLMs provide a flexible interface for jointly considering qualitative strategies, numerical order attributes, recent outcomes, and local task constraints. In delivery, an assigned order becomes a commitment that occupies a rider and removes a task from others' candidate observations. Studying this feedback requires the language-based decision policy to be coupled to shared assignment and routed execution. The question is how empirical personas are expressed through repeated, executable actions and which individual and collective patterns result.

We investigate this question with \emph{Large Language Models as Delivery Riders} (LLM-DR), a persona-grounded agentic simulation framework for on-demand delivery behavior. Empirical routed delivery waves define the persona profiles used in the agents' decision prompts. Each idle rider receives local offers, current state, short-term memory, and route-tool estimates to select rejection, single-order acceptance, or bundled acceptance. The environment validates order identifiers, resolves competing acceptances, and constructs the assigned pickup and drop-off sequence. Routing-tool calls supply road-network paths and travel times; the simulator executes assigned waves and updates location, availability, earnings, and memory. These updates carry the consequences of each action into later observations and the shared order pool. The delivery environment provides a common benchmarking task with fixed data inputs, interaction rules, and empirical evaluation procedures.\footnote{Research materials (code, de-identified data, and evaluation protocols): \url{https://github.com/Nicholas0027/LLM-DR}.} Persona-grounded LLM agents and rule-based policies provide reference implementations for comparing rider policies.

Three research questions guide the study of individual behavior and collective outcomes. First, which personas can be identified from empirical delivery waves, and how do their behavioral profiles differ? Second, do persona-grounded agents produce differentiated order actions and executed waves under a shared observed order stream? Third, how closely do persona-level outcomes resemble empirical behavior, and what fleet-level temporal and spatial patterns accumulate from the executed actions?

The main contributions are as follows.
\begin{itemize}
    \item We reveal empirical rider heterogeneity through unsupervised learning on routed delivery-wave trajectories. The resulting four personas capture systematic differences in work span, peak-hour preference, stacking tendency, route distance, and spatial footprint.
    \item We develop a persona-grounded agentic simulation methodology for on-demand delivery. Local order-action selection is coupled to shared-pool assignment, routing-tool calls, and state and memory updates, enabling controlled experiments on heterogeneous rider responses. The fixed order stream and evaluation protocol support reproducible benchmarking of rider policies.
    \item We connect rider-level behavioral validation with fleet-level temporal and spatial analysis. This cross-level evaluation identifies retained empirical persona orderings, displaced demand and service peaks, and shared central corridors alongside differentiated spatial reach.
\end{itemize}

\section{Related Work}
\subsection{On-Demand Delivery Simulation}
On-demand delivery models address platform coordination through simulation, recommendation, and assignment under changing demand. Multi-agent and reinforcement learning approaches have been used to study online food ordering and instant delivery strategies \cite{jinSimulationBasedScheduling2019,zouSimulationOnlineFood2023,zouOnlineFoodOrdering2022}. Rider-centered recommendation and recommending-and-grabbing approaches incorporate rider preferences into order allocation \cite{wangOnlineDeepReinforcement2023,wangRecommendingGrabbing2023}. Imitation-learning matching, predictive-reactive optimization, and graph-neural dispatch refine assignment and scheduling \cite{chenImitationLearningEnhanced2022,zhengPredictiveReactiveOptimization2023,chenOrderDispatchingGNN2024}. Such approaches evaluate rider-response representations through allocation objectives, including service efficiency and delivery reliability. Behavioral reconstruction adds a question about how those representations preserve differences in work timing, bundling, and spatial reach across repeated decisions. Empirical delivery reliability varies with real-time demand, restaurant density, and the allocation of different rider types \cite{liRealtimeDemandsRestaurant2025}. Route prediction studies also relate departures from shortest paths to perceived distance, deadlines, and individual preferences \cite{zhangRoutePredictionInstant2019}. These observations motivate empirical persona grounding and joint validation of selected orders, executed waves, and fleet-level outcomes.

\subsection{LLM Agents for Mobility and Agentic Social Simulation}
Mobility-focused LLM systems condition travel diaries or trajectories on personal profiles, activity patterns, and contextual motivations \cite{wangLargeLanguageModels2024,juTrajLLMModularLLMEnhanced2025,liBeMoreReal2024,shaoChainofPlannedBehaviourWorkflowElicits2024}. Their target output is typically an individual's activity sequence, with validation focused on temporal and spatial similarity. Generative agents extend the behavioral setting to interacting populations, using memory, planning, and reflection to support repeated social actions \cite{parkGenerativeAgentsInteractive2023}. Broader LLM-based simulation research examines collective behavior and the validation of artificial populations \cite{mouIndividualSocietySurvey2024,gaoLargeLanguageModelsEmpowered2024,adornettoGenerativeAgentsAgent2025}. Agent benchmarks further emphasize decision making, tool use, and behavior across multi-turn environments \cite{liuAgentBenchEvaluating2024,changSurveyEvaluationLLMs2024}. Delivery requires a specific coupling between local choices, exclusive order assignment, road-network commitments, and subsequent task availability. LLM-DR examines this coupling through repeated persona-conditioned decisions under a shared observed order stream.

\subsection{Agent Capabilities: Planning, Action, Tool Use, and \mbox{Memory}}
Agentic LLM systems connect task-oriented planning to consequential action in an external environment. ReAct-style agents interleave reasoning traces with actions, while tool-augmented models access external services for task-specific information \cite{yaoReActSynergizingReasoning2023,schickToolformerLanguageModels2023,qinToolLLMFacilitating2024}. Memory and reflection support repeated decisions, and explicit tool interfaces provide domain-specific calculations \cite{shinnReflexionLanguageAgents2023,zhangSurveyMemoryMechanism2025,goodellLargeLanguageModelAgents2025}. These capabilities connect data-driven representations to responsive agentic systems, including digital twins that integrate models, reasoning, and human interaction \cite{sanEvolutionDigitalTwins2026}. Transportation applications bring this interaction loop to task goals, changing local conditions, and spatial tools \cite{yuPreparingAgenticEra2025,yuAgenticVehicles2026}.

In LLM-DR, planning evaluates offered orders against persona-specific preferences, and action selection proposes rejection, single-order acceptance, or bundled acceptance. Routing-tool estimates inform that choice, while assigned actions trigger further tool calls for road-network execution. The environment constructs the waypoint sequence and processes execution; completed waves update state and short-term memory for later decisions. This allocation of responsibilities connects language-based rider decisions to shared platform feedback and executable mobility.

\section{Problem Formulation}
\subsection{Simulation Overview}
\begin{figure*}[!t]
\centering
\includegraphics[width=0.98\textwidth]{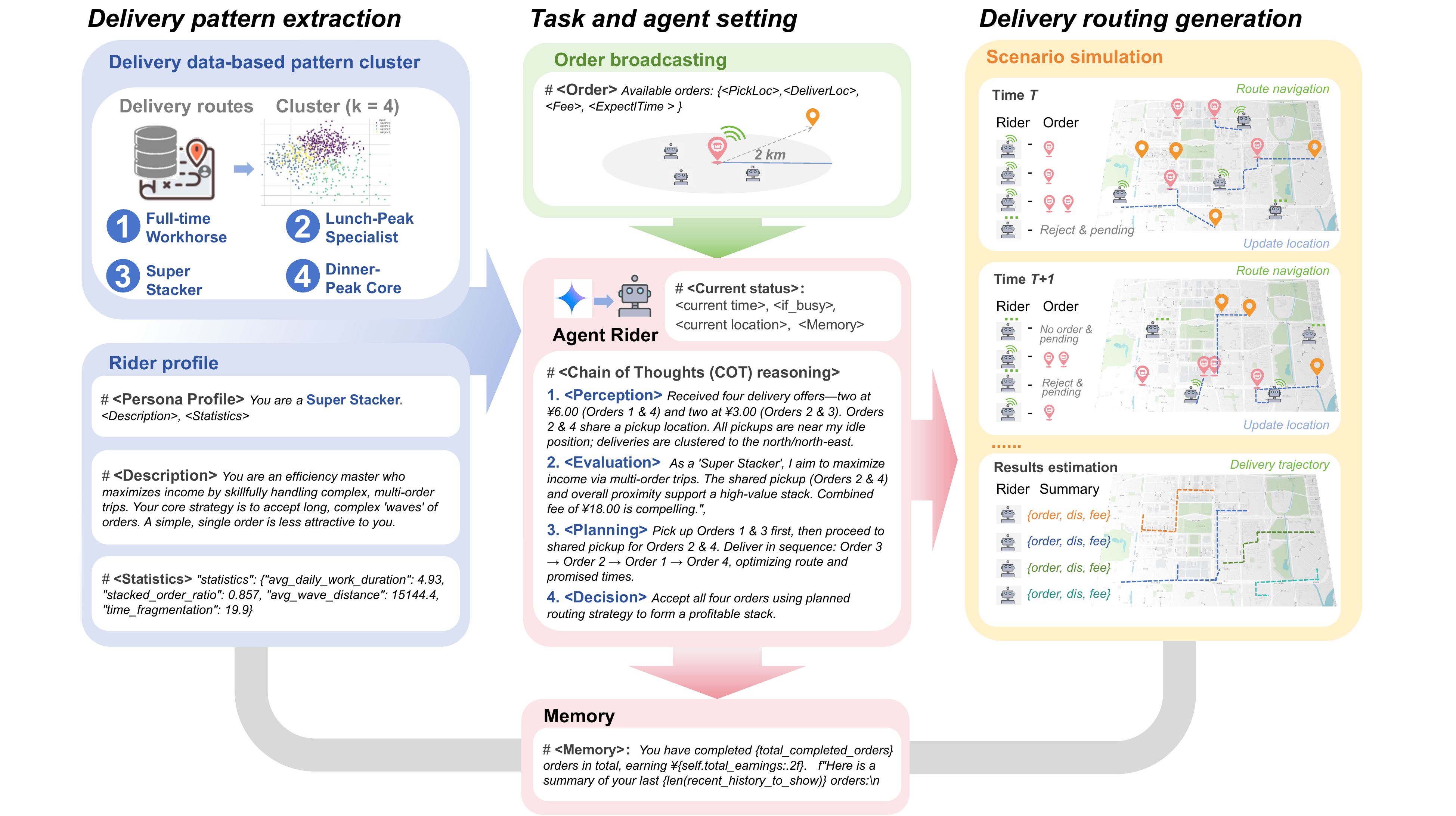}
\caption{Workflow of LLM-DR. Empirical trajectory data are converted into rider personas, instantiated as persona-grounded LLM agents, and evaluated through decision, tool-call, and route-execution actions in a dispatch simulation driven by real orders.}
\label{fig:workflow}
\end{figure*}

Fig.~\ref{fig:workflow} summarizes how empirical data, rider decisions, and platform execution form the simulation loop. Order and delivery-wave records provide behavioral features from which four personas are extracted. Each persona combines a strategy description with empirical group statistics and remains fixed during a run. The observed order stream supplies tasks to a shared pending pool, from which the platform exposes local offers to idle riders. An LLM agent considers its persona, current state, recent completed orders, and per-order route estimates to propose an order action.

The environment checks the proposed identifiers, resolves competing acceptances, and determines the realized assignment. It constructs a pickup-then-drop-off waypoint sequence, and the cycling-routing tool returns road-network paths, distances, and travel times. The simulator executes the assigned wave and updates rider location, availability, earnings, and memory when completion is processed. Assigned orders leave the pending pool, connecting each rider's commitments to other riders' subsequent opportunities. Action records support both empirical persona-level comparisons and analysis of fleet-level service rhythms, activity spaces, and shared road use. The definitions below formalize this observation--decision--assignment--execution loop and its resulting outcomes.

\subsection{System Definition and Dynamics}
We represent the delivery platform as a spatial, discrete-time multi-agent environment
\begin{equation}
\mathfrak{E}=\langle\mathcal{R},\mathcal{O},\mathcal{L},G,
\mathcal{Z},
\mathcal{D}_{\boldsymbol{\kappa}},
\mathcal{T}_{\boldsymbol{\kappa}},\Delta t\rangle .
\end{equation}
Here, $\mathcal{R}$ is the rider population, $\mathcal{O}$ is the exogenous order stream, $\mathcal{L}$ is the spatial domain and road network, $G$ is the cycling-routing interface, and $\mathcal{Z}$ maps system states to rider observations. The platform mechanism $\mathcal{D}_{\boldsymbol{\kappa}}$ controls offer exposure and conflict-resolved assignment, while $\mathcal{T}_{\boldsymbol{\kappa}}$ advances the environment by one interval $\Delta t$. Its fixed parameters are $\boldsymbol{\kappa}=(h,\rho,K,B,\tau_s)$: the pending-order horizon, broadcast radius, number of candidate riders per order, maximum offers retained per rider, and service time per order. Riders are the decision-making agents. Restaurants and customers enter through observed orders, and platform dispatch and routing operate as environmental mechanisms.

Each order $o\in\mathcal{O}$ has the fixed descriptor
\begin{equation}
u_o=(id_o,\ell_o^{p},\ell_o^{d},
\tau_o^{c},\tau_o^{ddl},f_o)
\end{equation}
and dynamic state
\begin{equation}
x_{o,t}=(q_{o,t},\alpha_{o,t},\tau_{o,t}^{f}),
\end{equation}
where $\ell_o^{p}$ and $\ell_o^{d}$ are the pickup and drop-off locations, $\tau_o^{c}$ is creation time, $\tau_o^{ddl}$ is the delivery deadline, and $f_o$ is the fee. All input orders are initialized in the pending pool and become eligible through their creation times. The status $q_{o,t}\in\{\textsc{pending},\textsc{assigned},\textsc{completed}\}$, $\alpha_{o,t}$ records the assigned rider when one exists, and $\tau_{o,t}^{f}$ is the simulated completion time.

Rider $r\in\mathcal{R}$ has a fixed persona $\psi_r$ and the dynamic state
\begin{equation}
x_{r,t}=(\ell_{r,t},s_{r,t},\mathcal{W}_{r,t},
\tau_{r,t}^{fin},e_{r,t},d_{r,t},H_{r,t}),
\end{equation}
where $\ell_{r,t}$ is location; $s_{r,t}\in\{\textsc{idle},\textsc{delivering}\}$ is operational status; $\mathcal{W}_{r,t}$ is the currently assigned wave; $\tau_{r,t}^{fin}$ is its scheduled finish time; $e_{r,t}$ and $d_{r,t}$ are cumulative earnings and routed distance; and $H_{r,t}$ is short-term delivery memory. For empirical rider $i$, let $z_i$ denote the standardized feature vector used for clustering. Persona $\psi$ combines a qualitative strategy description with the empirical prototype
\begin{equation}
\mu_{\psi}=\frac{1}{|\mathcal{R}_{\psi}^{emp}|}
\sum_{i\in\mathcal{R}_{\psi}^{emp}}z_i .
\end{equation}
The centroid summarizes the cluster in standardized feature space. The corresponding group statistics in their original units supply the persona prompt, together with its qualitative strategy description. These quantities remain fixed throughout the simulation.

The dynamic system state is
\begin{equation}
X_t=\left(t,\{x_{o,t}\}_{o\in\mathcal{O}},
\{x_{r,t}\}_{r\in\mathcal{R}}\right).
\end{equation}
Order descriptors, personas, spatial data, and environment parameters remain fixed during a run and are therefore omitted from $X_t$. The initial condition places an equal number of riders from each persona at locations sampled from the empirical pickup-density distribution $\nu_{pick}$:
\begin{equation}
\begin{aligned}
\ell_{r,t_0}&\sim\nu_{pick},\\
s_{r,t_0}&=\textsc{idle},\qquad \tau_{r,t_0}^{fin}=\emptyset,\\
(\mathcal{W}_{r,t_0},e_{r,t_0},d_{r,t_0},H_{r,t_0})
&=(\emptyset,0,0,\emptyset).
\end{aligned}
\end{equation}
The sampled locations receive the small spatial perturbation specified in the experimental design. At decision epoch $t$, the platform first forms the eligible pending pool
\begin{equation}
\mathcal{Q}_t=\left\{o:
\begin{aligned}
q_{o,t}&=\textsc{pending},\quad \tau_o^{ddl}>t,\\
t-h&\leq\tau_o^c<t+\Delta t
\end{aligned}\right\}.
\end{equation}
This pool uses interval batching: orders created during $[t,t+\Delta t)$ are exposed at the interval's start, together with eligible orders from the preceding $h$ minutes. Accepted waves also start at $t$. The convention gives the decision step access to arrivals within the upcoming interval; with $\Delta t=10$ minutes, its timing resolution is therefore one batch. Each order in $\mathcal{Q}_t$ is exposed to its $K$ nearest idle riders within radius $\rho$. For rider $r$, the resulting offers are ranked by deadline, rider-to-pickup distance, and fee, and the first $B$ define the candidate set
\begin{equation}
\mathcal{C}_{r,t}=
\operatorname{Top}_{B}^{\prec_{r,t}}\!\left(
\{o\in\mathcal{Q}_t:r\in N_K(o;\rho,X_t)\}
\right).
\end{equation}
Here, $N_K(o;\rho,X_t)$ is the set of up to $K$ idle riders nearest to $\ell_o^p$ within $\rho$, and $\prec_{r,t}$ orders offers by delivery deadline, pickup distance, and fee.

The routing interface returns road-network information between two waypoints:
\begin{equation}
G(\ell_i,\ell_j)=(d_{ij},\delta_{ij},p_{ij}),
\end{equation}
where $d_{ij}$, $\delta_{ij}$, and $p_{ij}$ are cycling distance, estimated travel time, and route polyline. Before action selection, the environment constructs
\begin{equation}
E_{r,t}=\left\{
G(\ell_{r,t},\ell_o^p),G(\ell_o^p,\ell_o^d)
:o\in\mathcal{C}_{r,t}\right\}.
\end{equation}
The LLM does not observe $X_t$ directly. The observation map gives
\begin{equation}
y_{r,t}=\mathcal{Z}_r(X_t)=\left(t,\ell_{r,t},e_{r,t},d_{r,t},H_{r,t},
\{u_o:o\in\mathcal{C}_{r,t}\},E_{r,t}\right).
\end{equation}

An idle rider proposes a bid $b_{r,t}\in2^{\mathcal{C}_{r,t}}$. The empty set denotes voluntary rejection, while a set with two or more orders denotes proposed bundling. LLM-DR implements the persona-grounded policy as
\begin{align}
\widetilde m_{r,t}&\sim
p_{\theta}\!\left(\cdot\mid P(y_{r,t},\psi_r)\right),\\
b_{r,t}&=\mathcal{V}(\widetilde m_{r,t};\mathcal{C}_{r,t}),
\end{align}
where $p_{\theta}$ is the response distribution induced by LLM $M_{\theta}$, $P$ is the structured prompt, and $\mathcal{V}$ parses response $\widetilde m_{r,t}$, removes duplicate or unavailable identifiers, and yields a valid subset of the candidate set. Validation checks order identifiers and availability. Deadline considerations enter through the prompt and route estimates; $\mathcal{V}$ applies no joint-route deadline or carrying-capacity test. Several riders may bid for the same order, so platform conflict resolution converts these proposals into assignments.

Let $\pi_t$ order all nonempty bids by increasing mean rider-to-pickup distance, with rider ID breaking ties. The platform converts bids into pairwise-disjoint realized assignments according to
\begin{equation}
a_{\pi_t(j),t}=b_{\pi_t(j),t}\setminus
\bigcup_{i<j}a_{\pi_t(i),t}.
\end{equation}
Thus, $a_{r,t}$ is the realized rider action after platform conflict resolution; it can be empty even when $b_{r,t}$ was nonempty. Removing assigned orders from the pending pool couples rider outcomes through $\mathcal{D}_{\boldsymbol{\kappa}}$.

For a nonempty assignment $a_{r,t}$, let $(o_1,\ldots,o_k)$ retain the environment's candidate ranking. The executed wave follows the waypoint chain
\begin{equation}
\chi_{r,t}=\left(\ell_{r,t},\ell_{o_1}^{p},\ldots,
\ell_{o_k}^{p},\ell_{o_1}^{d},\ldots,\ell_{o_k}^{d}\right).
\end{equation}
If consecutive waypoints in $\chi_{r,t}$ are indexed by $(v_j,v_{j+1})$, routed distance and duration are
\begin{align}
D(\chi_{r,t})&=\sum_j d_{v_jv_{j+1}},\\
T(\chi_{r,t})&=\sum_j\delta_{v_jv_{j+1}}+k\tau_s.
\end{align}
The routing tool therefore makes an accepted wave spatially executable, while waypoint construction remains an environment rule. The rider becomes \textsc{delivering} with scheduled finish time $\tau_{r,t}^{fin}=t+T(\chi_{r,t})$ and is released at the first decision epoch at or after that time. All orders in the wave receive the common completion timestamp $\tau_{o,t}^{f}=\tau_{r,t}^{fin}$. This timestamp is used to evaluate a wave-end on-time proxy against each order's deadline. The recorded action chain preserves the pickup and drop-off sequence, while operational availability and completion statistics are updated at wave level.

Let $\mathbf a_t=(a_{r,t})_{r\in\mathcal{R}}$ and define the orders released during the current interval as
\begin{equation}
\mathcal{O}^{arr}_{[t,t+\Delta t)}=
\left\{o\in\mathcal{O}:t\leq\tau_o^c<t+\Delta t\right\}.
\end{equation}
The state transition is
\begin{equation}
X_{t+\Delta t}=\mathcal{T}_{\boldsymbol{\kappa}}
\left(X_t,\mathbf a_t,
\mathcal{O}^{arr}_{[t,t+\Delta t)},G\right).
\end{equation}
At each decision epoch, the simulator first completes waves whose scheduled finish times have been reached and releases their riders. It then exposes the interval-batched pool, collects bids, applies assignments, and advances the clock. Completion processing moves a rider to the final drop-off and increments earnings and mileage. Memory is updated as
\begin{equation}
H_{r,t'}=\mathcal{U}
\left(H_{r,t},a_{r,t},
e_{r,t'}-e_{r,t},D(\chi_{r,t}),\mathbf c_{r,t}\right),
\end{equation}
where $\mathbf c_{r,t}$ records completion and wave-end deadline outcomes and $t'$ is the decision epoch at which the completed wave is processed. The scheduled wave finish time is retained in the action log. Rider interaction occurs through bids that alter the shared pending pool and hence one another's subsequent observations.

Let $t_H$ be the end of the order-release window. After the last decision epoch, the environment completes all remaining assigned waves and reaches terminal time $t_T\geq t_H$. Let $\mathcal{A}_r$ contain the rider's decision records and completed waves over the full run. Rider-level operational outcomes are
\begin{equation}
Y_r=g(x_{r,t_T},\mathcal{A}_r),
\end{equation}
where $g$ returns the delivery, earnings, bundling, mileage, and deadline indicators used in the evaluation. These aggregates use the full action record, while the decision prompt receives short-term memory. Empirical similarity is reported as a metric vector:
\begin{equation}
\mathbf D_{\psi}=\left[
D_m\!\left(\mathbb{P}^{sim}_{\psi,m},
\mathbb{P}^{emp}_{\psi,m}\right)
\right]_{m\in\mathcal{M}}.
\end{equation}
Here, $\mathcal{M}$ contains the temporal, spatial, routed-distance, and bundle-structure measures. Table~\ref{tab:validation_results} reports the order-time, spatial, and distance components; Fig.~\ref{fig:persona_validation} adds rider-level comparisons of volume, bundling, work span, and spatial footprint across persona groups.

\section{LLM-DR Agentic Simulation Framework}
\subsection{Empirical Persona Extraction}
The first stage extracts rider personas from real delivery trajectories. We use 9,025 filtered order records and a reconstructed sample of 2,508 routed delivery waves associated with 710 couriers. Observed pickup and delivery action sequences provide the wave structure; cycling routes connect consecutive action locations. For each rider, we calculate 15 behavioral features covering temporal preference, operating intensity, order stacking, spatial footprint, and platform attributes. Examples include daily work duration, lunch and dinner order ratios, active 10-minute bins, average wave distance, average wave duration, simultaneous-load ratio, activity-area size, rider level, speed, and maximum load. Order-based indicators use the order records, and wave-based indicators use the reconstructed wave sample.

We apply $\log(1+x)$ to skewed continuous variables and standardize every feature to zero mean and unit variance. Clustering uses $k$-means with $k$-means++ initialization and a squared-Euclidean objective. Each fitted solution retains the lowest-inertia result from 80 initializations, with a maximum of 300 iterations and a center-displacement tolerance of $10^{-5}$. We evaluate $k=2$ to $8$ with inertia, silhouette score, Calinski-Harabasz score, and Davies-Bouldin score. We select $k=4$ for its parsimonious, interpretable behavioral partition. The four-cluster solution has a silhouette score of 0.108 and a mean adjusted Rand index of 0.927 across 40 stability runs. The clusters thus summarize overlapping behavioral profiles with reproducible membership; the persona names describe their relative feature patterns.

\begin{figure*}[!t]
\centering
\includegraphics[width=0.98\textwidth]{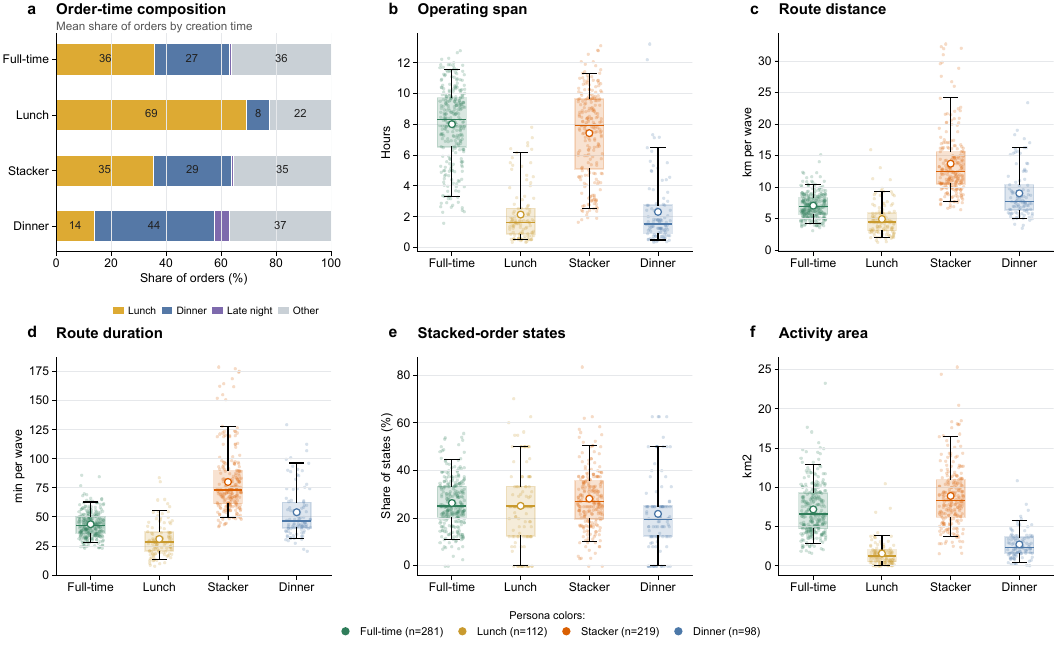}
\caption{Empirical delivery-pattern fingerprints of the four rider personas. Panel a reports mean order-time composition. Panels b--f report rider-level distributions of operating span, routed distance, routed duration, stacked-order states, and activity area. Points denote individual riders, boxes show interquartile ranges, and whiskers show the 5th--95th percentiles.}
\label{fig:persona_patterns}
\end{figure*}

The resulting personas are:
\begin{itemize}
    \item \textbf{Full-time Workhorse}: long work span, many active time bins, and stable order volume.
    \item \textbf{Lunch-Peak Specialist}: concentrated activity during 11:00--14:00, short delivery waves, and compact spatial coverage.
    \item \textbf{Super Stacker}: long routed waves, large spatial footprint, and strong preference for multi-order trips.
    \item \textbf{Dinner-Peak Core}: concentrated evening activity and simpler delivery waves.
\end{itemize}
Fig.~\ref{fig:persona_patterns} shows the empirical distributions behind these personas.

\subsection{Dispatch Environment}
The simulation environment ingests real order data and advances in 10-minute intervals. At each interval's start, the platform exposes orders created within that interval and eligible pending orders created during the preceding 30 minutes, following the batching convention in (8). It finds idle candidate riders within a 2-km dispatch radius and offers each order to at most three nearby riders. Each rider retains up to six offers ranked by deadline, pickup distance, and fee. After riders make decisions, the platform prioritizes accepted bids with shorter mean rider-to-pickup distances and assigns the still-available orders in that priority order. Orders not assigned remain pending for later intervals while satisfying the visibility-window and deadline conditions. Riders carrying out an assigned wave receive their next offers after that wave has finished.

This design approximates platform order pooling and local rider competition while keeping the mechanism transparent. It also avoids assuming that every rider observes every order in the system. Initial rider locations are sampled from empirical pickup locations in the evaluation window, with small spatial perturbations, so that the synthetic fleet follows observed demand density.

\subsection{Persona-Grounded LLM Decision Protocol}
For each idle rider, LLM-DR constructs a structured prompt from the local observation in (12) and the rider persona. It contains the current location, accumulated earnings, recent delivery memory, current time, candidate-order attributes, and route-tool estimates from the current location to pickup and from pickup to drop-off. The prompt asks the agent to evaluate orders using persona-specific goals, including time-window preference, income stability, order-stacking tendency, distance tolerance, and anticipated deadline pressure. The model returns a JSON response containing accepted order IDs and a concise rationale. The action-validity operator checks the proposed identifiers; the dispatch environment then resolves conflicts and produces the realized assignment used for routing and execution.

The agent receives only its local candidate batch, with up to five recently completed orders in memory. The full-day order list and empirical rider action sequences are absent from the decision prompt. An accepted-ID list specifies the response to the offered set, with an empty list denoting rejection. All personas and baselines share this information boundary and dispatch convention.

The main agent capabilities are tied to distinct simulation roles. The persona profile supplies an empirical behavioral prior for heterogeneous time use, distance tolerance, and stacking preference. The routing tool supplies road-network distance, travel time, and route geometry before each decision and after each accepted wave. The short-term memory records completed orders, earnings, mileage, and recent delivery outcomes, which lets the prompt represent accumulated work state. The structured JSON protocol makes the accept, bundle, or reject action auditable and comparable across agents.

\begin{figure*}[!t]
\centering
\includegraphics[width=0.98\textwidth]{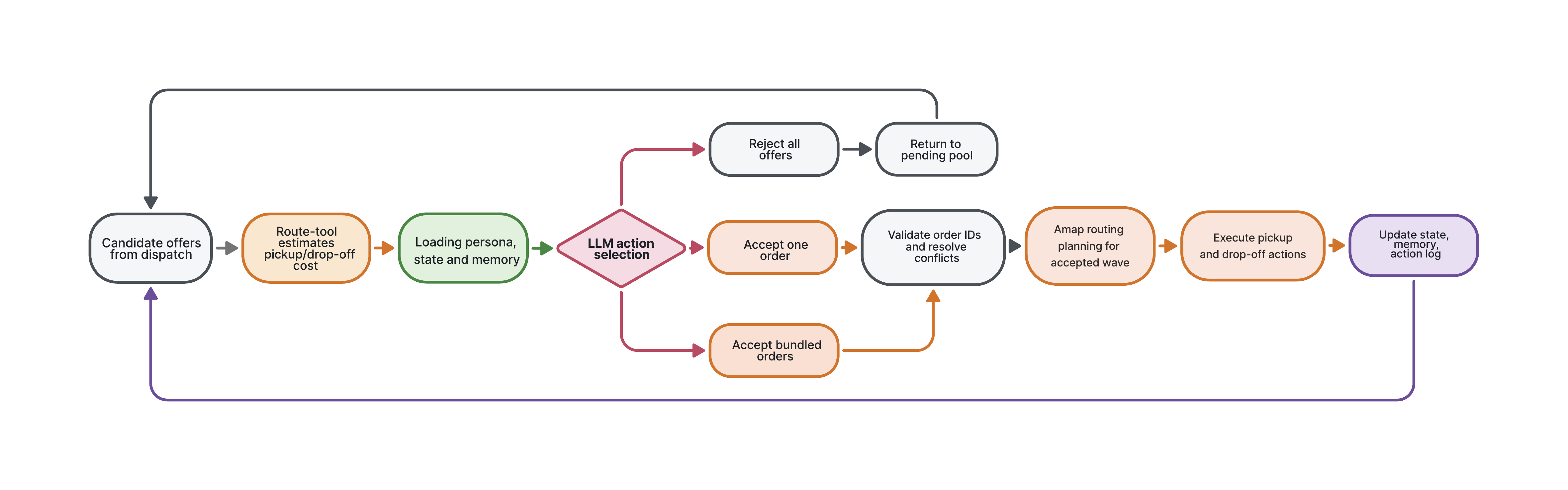}
\caption{Action chain decision tree for one rider decision step. The process moves from candidate offers to route-tool estimation, prompt construction, LLM action selection, environment validation, route execution, and memory/action-log update. The rejection branch returns orders to the pending pool, while accepted single-order or bundled waves proceed through conflict resolution and route execution.}
\label{fig:action_chain_tree}
\end{figure*}

Fig.~\ref{fig:action_chain_tree} unpacks one rider decision step as an action chain. Candidate offers first trigger route-tool estimates. The prompt then combines persona, memory, order attributes, and route estimates before the LLM proposes rejection, single-order acceptance, or bundled acceptance. The environment validates the selected order IDs, resolves conflicts, calls the routing tool for realized waves, executes pickup and drop-off actions, and writes the updated state to memory and the action log.

\subsection{Routing Tool and Action Chain Update}
If a rider receives a nonempty realized assignment, LLM-DR constructs the delivery-wave waypoint sequence from the rider's current location, the assigned pickup points, and the assigned drop-off points, using the fixed pickup-then-drop-off ordering defined in (16). For a two-order assignment, the chain is current location $\rightarrow$ pickup$_1$ $\rightarrow$ pickup$_2$ $\rightarrow$ drop-off$_1$ $\rightarrow$ drop-off$_2$. Each consecutive pair invokes the Amap cycling-routing API; segment distances, travel times, and polylines are cached locally and aggregated into a wave-level route. The prompt's pre-decision estimates describe each offered order separately, while this complete chain is constructed after assignment. A fixed 120-second service time per order is added to the routed travel time to represent pickup and handoff operations. Orders in the same wave share its end timestamp, and rider status, location, earnings, and memory are updated when the completed wave is processed. The fixed waypoint order and wave-level completion convention apply to every compared method.

Algorithm~\ref{alg:llmdr} summarizes the executable persona-grounded simulation procedure.

\begin{algorithm}[!t]
\caption{Persona-Grounded Agentic Delivery Simulation in LLM-DR}
\label{alg:llmdr}
\footnotesize
\begin{algorithmic}[1]
\STATE \textbf{Input:} empirical orders $\mathcal{O}$; persona profiles $\Psi$; riders per persona $n$; horizon $t_H$; settings $\boldsymbol{\kappa}=(h,\rho,K,B,\tau_s)$ and $\Delta t$; routing tool $G$; LLM $M_\theta$.
\STATE \textbf{Output:} executed action chains $\mathcal{A}$, rider outcomes $\mathcal{Y}$, and persona-level validation components $\mathbf{D}$.
\STATE Initialize $n$ \textsc{idle} rider agents per $\psi\in\Psi$ at pickup-density-sampled locations, with profile $\psi$, memory $H_r$, pending pool, $\mathcal{A}\leftarrow\emptyset$, and $t\leftarrow t_0$.
\WHILE{$t\leq t_H$}
    \STATE Complete finished waves and update $x_{r,t}$; form interval-batched pool $\mathcal{Q}_t$ using (8).
    \FOR{each $o\in\mathcal{Q}_t$}
        \STATE Offer $o$ to its $K$ nearest \textsc{idle} riders within radius $\rho$, yielding candidate sets $\mathcal{C}_{r,t}$.
    \ENDFOR
    \STATE $\mathcal{B}_t\leftarrow\emptyset$.
    \FOR{each \textsc{idle} rider $r$ with $\mathcal{C}_{r,t}\neq\emptyset$}
        \STATE Retain up to $B$ offers; obtain $E_{r,t}$ from $G$ and construct local observation $y_{r,t}$.
        \STATE $\widetilde m_{r,t}\leftarrow M_\theta(P(y_{r,t},\psi_r))$ and $b_{r,t}\leftarrow\mathcal{V}(\widetilde m_{r,t};\mathcal{C}_{r,t})$; retry once if JSON is unparseable.
        \STATE Append the proposed action to $\mathcal{A}$; if $b_{r,t}\neq\emptyset$, add $(r,b_{r,t})$ to $\mathcal{B}_t$.
    \ENDFOR
    \STATE Sort $\mathcal{B}_t$ by mean rider-to-pickup distance, then rider ID.
    \FOR{each bid $(r,b)\in\mathcal{B}_t$ in priority order}
        \STATE $a_{r,t}\leftarrow\{o\in b\mid o\text{ is still pending and unassigned}\}$; \textbf{continue} if $a_{r,t}=\emptyset$.
        \STATE Construct and route $\chi_{r,t}$ with $G$; assign $a_{r,t}$ to $r$, set common finish time $t+T(\chi_{r,t})$, mark $r$ \textsc{delivering}, and log the wave in $\mathcal{A}$.
    \ENDFOR
    \STATE Advance $t\leftarrow t+\Delta t$.
\ENDWHILE
\STATE Complete remaining assigned waves to reach $t_T$; aggregate $\mathcal{Y}$ and compute the validation vector $\mathbf{D}$ against empirical behavior by persona.
\end{algorithmic}
\end{algorithm}

\section{Experimental Design}
\subsection{Dataset and Evaluation Window}
The experiment uses a real-world Beijing delivery order dataset from February 15, 2020 \cite{zhangDecipheringDeliveryMobility2025,zhangUncoveringCommunityStructure2026,zhangVisualizingSpatiotemporalFlows2026}. The full file contains 9,025 distinct orders. The main evaluation uses 8,656 orders created between 08:00 and 20:00, which covers the primary daytime service window. These records involve 705 distinct observed couriers during the window; this count describes rider participation across the day rather than simultaneous online supply. The order records provide the order-time, pickup-location, and origin--destination reference distributions. Wave-level comparisons use the reconstructed waves starting within the evaluation window.

The simulator assigns an experimental distance-based reward: CNY 3.0 for orders up to 1 km, CNY 6.0 for 1--3 km, CNY 10.0 for 3--5 km, and CNY 15.0 for orders longer than 5 km. The brackets use straight-line pickup-to-drop-off distance. This assumed reward schedule supplies a common decision input for every method; reported earnings are simulated rewards under that schedule. Road-network distance from the routing tool is used separately for travel estimates and execution.

We instantiate a 60-rider fleet with 15 riders per persona, giving each profile equal representation for rider-side comparison. The observed order stream remains unchanged, producing a controlled fleet configuration with approximately 144 input orders per agent, compared with 12 orders per distinct observed courier over the window. These ratios describe window-level workload and participation; concurrent demand and availability also vary over time and space. The main agentic experiment has no global rider-decision cap and no per-step rider-decision cap. Every idle rider who receives at least one candidate order can make a decision. Decision counts vary across policies because their choices change availability, bundle size, and later dispatch opportunities.

This is a conditional single-day reconstructive validation. The same service day supplies the empirical personas, order stream, and reference behavioral distributions. The persona statistics are fixed before the run, and the LLM acts on local offers without receiving observed rider action sequences. Validation measures how those data-derived profiles carry through to executed actions under the specified fleet and dispatch conditions. Agreement in persona-defining indicators therefore measures reconstructive consistency, with order-time, pickup-grid, and route-distance discrepancies reported alongside it.

Benchmark comparisons retain the order stream, persona composition, initialization seed, local observations, dispatch rules, reward schedule, routing interface, and completion conventions. Alternative policies replace the rider's order-selection rule. Common validation and analysis procedures assess delivery performance, empirical behavioral discrepancies, and collective temporal and spatial outcomes.

\subsection{LLM and Routing Implementation}
The main LLM-DR run uses DeepSeek-v4-pro through an OpenAI-compatible chat-completions interface. The request temperature is 0.2, the maximum output length is 1024 tokens, and JSON response mode is enabled when supported by the endpoint. Each decision prompt is submitted with a system instruction requiring strict JSON output. The simulator extracts the first JSON object from the response, validates the accepted order IDs against the offered candidate set, and treats invalid decisions as rejection after one retry. The main run contains 1,587 rider decisions.

All routed distances, travel times, and polylines are generated by the Amap bicycling-routing API. Route estimates used inside prompts and accepted-wave action chains share the same routing interface. Requests are cached locally, with 5,002 route requests and 10,253 route-cache hits in the main run. No Amap routing failure was recorded in the final main experiment.

\subsection{Baselines}
We compare LLM-DR with four deterministic behavioral references under the same order stream, dispatch radius, initialization, Amap cycling-routing tool, and conflict-resolution rule:
\begin{itemize}
    \item \textbf{Nearest order}: accepts the geographically nearest candidate order.
    \item \textbf{Highest fee}: accepts the candidate order with the highest immediate reward.
    \item \textbf{Fee-distance optimization}: accepts orders that maximize reward per estimated travel distance.
    \item \textbf{Persona heuristic}: applies fixed persona-specific rules for time-window preference and stacking.
\end{itemize}

The four baselines provide transparent behavioral references. Nearest order and highest fee select one order, so their stacked ratio is zero by construction. Learned dispatch methods generally require a separate training setup, proprietary assignment labels, or an explicit platform-efficiency objective. We restrict the comparison to behavioral rules so that all methods operate under the same rider-side decision and dispatch conditions.

\subsection{Evaluation Metrics}
We separate delivery performance from action chain validation. Delivery-performance metrics are averaged across riders, including decisions, completed orders, simulated earnings, average bundle size, and stacked-wave ratio. The on-time indicator for an order compares its wave-end timestamp with its delivery deadline; each rider's rate is averaged over completed orders, and Table~\ref{tab:main_results} reports the mean rider rate. This wave-end proxy can register an early drop-off as late when another order extends the wave. Avg. Stacked Ratio is the rider-level share of waves containing at least two orders.

Action-chain validation compares completed simulated trips with empirical rider behavior within each persona. Computational social simulation should preserve explanatory structure as well as operational performance, and human mobility has well-documented temporal and spatial regularities \cite{hofmanIntegratingExplanationPrediction2021,gonzalezUnderstandingIndividualHuman2008,songLimitsPredictabilityHuman2010}. Table~\ref{tab:validation_results} reports Jensen--Shannon divergence (JSD) for the hourly creation-time composition of served orders and for pickup grids; Wasserstein-1 (W1) and Kolmogorov--Smirnov (KS) distances for straight-line origin--destination and routed wave distances; absolute error in pickup radius of gyration; and log-ratio error of pickup activity area. The hourly statistic describes the demand selected by a persona. Execution timing is shown separately in Fig.~\ref{fig:temporal_outcomes}. Pickup radius of gyration and activity area in the table pool all completed-order pickup points within a persona, measuring group-level spatial reach. Lower distances or errors indicate closer agreement with the corresponding empirical reference.

Fig.~\ref{fig:persona_validation} compares rider-level distributions over the same evaluation window. Order volume uses the order records; mean orders per wave, stacked-state intensity, routed wave distance, and work span use each rider's available wave sample. Work span is the interval from the first wave start to the last wave end. Stacked-state intensity is the mean share of pickup/drop-off action states carrying more than one order, averaged across waves. Pickup activity area is the convex hull of each rider's pickup locations in both sources. For each metric and source, valid riders from all four personas are ranked together. If $R_i$ is rider $i$'s average rank, including tied values, and $N$ is the number of valid riders in that source, the plotted percentile is $p_i=(R_i-0.5)/N$. This empirical-rank transformation uses no histogram bins. Violin densities use Scott's bandwidth rule, boxes show the interquartile range, and whiskers mark the 10th and 90th percentiles. The panels compare within-source relative positions, complementing the untransformed discrepancy measures in Table~\ref{tab:validation_results}.

\section{Results}
\subsection{Delivery Performance Across Baselines and Personas}
The results are organized to answer the three research questions. Fig.~\ref{fig:persona_patterns} addresses whether empirical personas capture distinct rider profiles. Table~\ref{tab:main_results} examines whether persona-grounded agents produce differentiated operational behavior under the same order stream. Table~\ref{tab:validation_results} and Fig.~\ref{fig:persona_validation} then evaluate whether the simulated patterns resemble empirical ground-truth behavior. Figs.~\ref{fig:temporal_outcomes}, \ref{fig:spatial_outcomes}, and \ref{fig:trajectory3d} examine how those actions accumulate into fleet-level temporal and spatial outcomes.

Table~\ref{tab:main_results} reports delivery-performance profiles for the deterministic references and the four LLM-DR persona groups. The deterministic references are summarized at the 60-rider fleet level, whereas LLM-DR is reported through its four persona groups, with 15 agents in each group. Across the LLM-DR fleet, 1,428 of the 8,656 input orders are completed (16.5\%), and 7,228 remain unassigned. Completion coverage contextualizes the persona-level outcomes: they describe realized service under the specified supply and offer-exposure rules. Unassigned orders combine lack of rider exposure, rejected offers, assignment competition, and orders that expire from the pending window.

Shared-pool assignment also changed the order actions that riders could realize. Of 1,006 nonempty acceptance proposals, 31 received only a subset of the selected orders and 60 received none after conflict resolution. The realized assignment therefore differed from the proposal in 91 cases (9.0\%). This difference records competition through shared orders, linking individual action selection to platform-mediated service outcomes.

Table~\ref{tab:main_results} reveals heterogeneity in both action selection and action composition. Lunch-Peak Specialist agents encounter the most decision opportunities (42.067 per rider) and complete 16 orders. Their low output alongside frequent decisions is consistent with selective acceptance. Super Stacker agents show the opposite configuration: 16.600 decisions yield 30 completed orders, with a mean bundle size of 2.212 and a mean stacked-wave ratio of 0.813. Bundled actions therefore support more completed orders per decision opportunity. Full-time Workhorse agents complete 26 orders with average simulated earnings of CNY 137.800 and a bundle size of 1.523. Dinner-Peak Core agents combine 23 completed orders with the highest persona-level on-time proxy (0.861) and the simplest wave composition (1.116 orders per wave).

The deterministic references anchor these profiles. Nearest-order and highest-fee rules execute single-order waves, with on-time proxies of 0.877 and 0.860. Fee-distance optimization combines 25 completed orders with a bundle size of 1.697 and an on-time proxy of 0.568. Within LLM-DR, the Super Stacker profile produces more orders and rewards together with a lower wave-end on-time proxy. The interpretation follows the execution convention: larger waves extend the common completion timestamp used for every constituent order. Together, these indicators describe policy tradeoffs in realized workload and wave composition under the common environment. The empirical comparisons below assess how the resulting action patterns resemble recorded rider behavior.

\begin{table*}[!t]
\caption{Delivery Performance of Baselines and LLM-DR Persona Groups}
\label{tab:main_results}
\centering
\scriptsize
\setlength{\tabcolsep}{3pt}
\begin{tabular*}{0.70\textwidth}{@{\extracolsep{\fill}}lrrrrrr@{}}
\toprule
\textbf{Model / Persona} & \textbf{Avg. Dec.} & \textbf{Avg. Orders} & \textbf{Avg. Earn.} & \textbf{Avg. On-Time} & \textbf{Avg. Bundle} & \textbf{Avg. Stacked} \\
 &  &  & \textbf{(CNY)} & \textbf{Proxy} & \textbf{Size} & \textbf{Ratio} \\
\midrule
Nearest order & 26.900 & 24 & 114.600 & 0.877 & 1.000 & 0.000 \\
Highest fee & 22.883 & 20 & 117.450 & 0.860 & 1.000 & 0.000 \\
Fee-distance optimization & 16.033 & 25 & 128.950 & 0.568 & 1.697 & 0.697 \\
Persona heuristic & 27.317 & 17 & 100.633 & 0.659 & 1.374 & 0.374 \\
\midrule
LLM-DR(Full-time Workhorse) & 20.800 & 26 & 137.800 & 0.657 & 1.523 & 0.523 \\
LLM-DR(Lunch-Peak Specialist) & 42.067 & 16 & 78.467 & 0.837 & 1.271 & 0.267 \\
LLM-DR(Super Stacker) & 16.600 & 30 & 151.267 & 0.407 & 2.212 & 0.813 \\
LLM-DR(Dinner-Peak Core) & 26.333 & 23 & 112.733 & 0.861 & 1.116 & 0.116 \\
\bottomrule
\end{tabular*}
\par\vspace{1.5mm}
\begin{minipage}{0.70\textwidth}
\scriptsize
\raggedright
\emph{Note:} Values are per-rider averages over 60 riders for baselines and 15 agents for each LLM-DR persona. Avg. Orders is rounded to the nearest whole order. Earnings use the assumed reward schedule. On-Time Proxy compares the common wave-end timestamp with each order's deadline. Avg. Stacked Ratio is the share of multi-order waves per rider.
\end{minipage}
\end{table*}

\subsection{Action Chain Empirical Similarity}
Table~\ref{tab:validation_results} summarizes action chain empirical similarity for the four LLM-DR persona groups. It complements the delivery-performance indicators in Table~\ref{tab:main_results} by comparing served-order time composition, pickup geography, origin--destination distance, routed wave distance, and group-level spatial footprint. All indicators are distances or errors from the empirical distribution, so lower values indicate closer similarity.

The component metrics reveal distinct aspects of empirical similarity. Full-time Workhorse has the closest served-order time composition (JSD 0.015), together with small origin--destination distance errors. These values describe the time and spatial separation of selected demand. Lunch-Peak Specialist has the smallest routed wave-distance W1 value (0.764 km), while its order-time JSD is the largest among the four groups. Super Stacker combines a small order-time JSD (0.025) and the smallest origin--destination KS distance (0.093) with a larger routed wave-distance W1 value (4.158 km). Matching the distances of individual orders can therefore coexist with a discrepancy in the multi-stop routes formed from them. Dinner-Peak Core has the smallest pooled pickup radius-of-gyration error (1.290 km) and area error (0.137), but the largest wave KS distance (0.608). Its group-level spatial reach and wave geometry show different degrees of agreement.

Pickup-grid JSD varies from 0.230 to 0.245 across personas. All groups select orders from the same demand surface and use the same density-based initialization and local exposure rule, giving those spatial discrepancies a shared environmental context. Order-time composition, wave geometry, and pickup footprint provide additional persona-specific information. The component-wise evaluation separates demand selection from bundled execution and distinguishes group-level spatial reach from individual rider distributions.

\begin{table*}[!t]
\caption{Action Chain Validation Against Empirical Rider Behavior}
\label{tab:validation_results}
\centering
\scriptsize
\setlength{\tabcolsep}{2pt}
\begin{tabular*}{0.70\textwidth}{@{\extracolsep{\fill}}lrrrrrrrr@{}}
\toprule
\textbf{LLM-DR Persona} & \textbf{Order-Time} & \textbf{Pickup} & \textbf{OD W1} & \textbf{OD KS} & \textbf{Wave W1} & \textbf{Wave KS} & \textbf{$R_g$ Err.} & \textbf{Area Err.} \\
 & \textbf{JSD} & \textbf{JSD} & \textbf{(km)} &  & \textbf{(km)} &  & \textbf{(km)} & \textbf{Log} \\
\midrule
Full-time Workhorse & 0.015 & 0.230 & 0.126 & 0.096 & 1.330 & 0.148 & 3.221 & 1.496 \\
Lunch-Peak Specialist & 0.129 & 0.239 & 0.114 & 0.121 & 0.764 & 0.175 & 6.253 & 1.799 \\
Super Stacker & 0.025 & 0.233 & 0.162 & 0.093 & 4.158 & 0.284 & 6.448 & 1.551 \\
Dinner-Peak Core & 0.064 & 0.245 & 0.269 & 0.132 & 4.516 & 0.608 & 1.290 & 0.137 \\
\bottomrule
\end{tabular*}
\par\vspace{1.5mm}
\begin{minipage}{0.70\textwidth}
\scriptsize
\raggedright
\emph{Note:} Lower values indicate closer agreement with the empirical reference. JSD denotes Jensen--Shannon divergence; W1 denotes Wasserstein-1 distance; KS denotes Kolmogorov--Smirnov distance. Order-Time JSD uses served-order creation times. $R_g$ and Area errors use pickup points pooled within each persona; wave metrics use the reconstructed wave sample.
\end{minipage}
\end{table*}

\begin{figure*}[!t]
\centering
\includegraphics[width=0.92\textwidth]{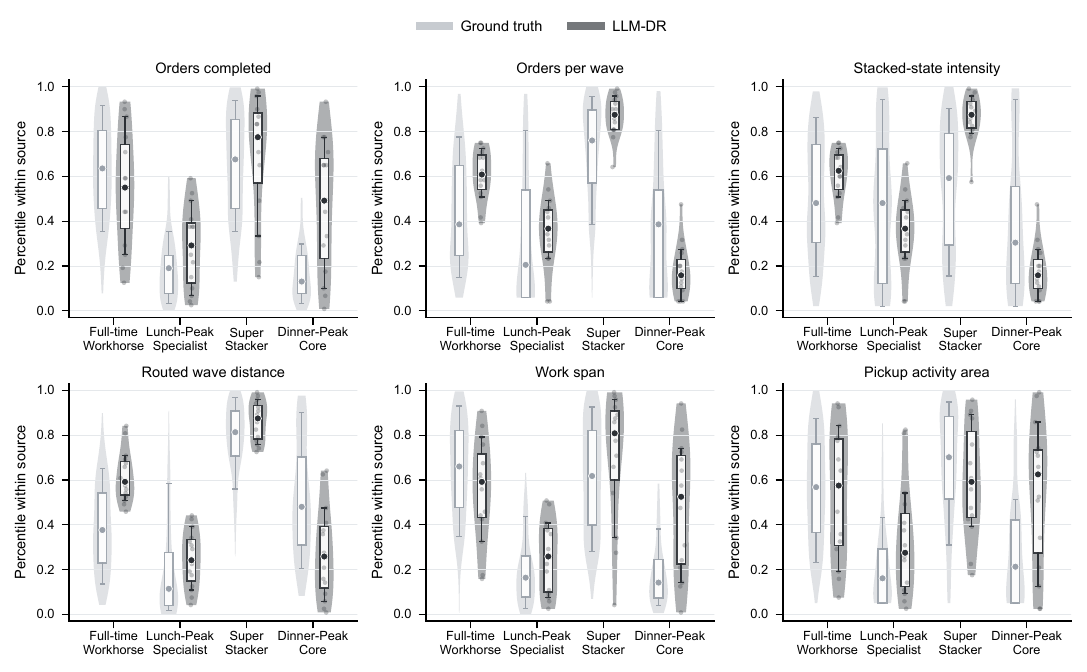}
\caption{Ground-truth and simulated rider distributions over 08:00--20:00, expressed as within-source average-rank percentiles across the four personas. Ground truth includes 281 Full-time Workhorse, 111 Lunch-Peak Specialist, 219 Super Stacker, and 94 Dinner-Peak Core riders; simulation includes 15 per persona. Violins use Scott's bandwidth; boxes, whiskers, and points indicate interquartile ranges, 10th--90th percentiles, and medians. Simulated riders are overlaid. Wave indicators use the reconstructed wave sample; pickup area uses pickup-only convex hulls. Super Stacker retains the highest median ranks in order volume, bundling intensity, and routed-wave distance.}
\label{fig:persona_validation}
\end{figure*}

Fig.~\ref{fig:persona_validation} examines the relative positions of individual riders within each source population. Super Stacker has the highest persona median for completed orders, orders per wave, stacked-state intensity, and routed wave distance in both sources. For routed wave distance, its median percentile is 0.813 in ground truth and 0.875 in simulation; Lunch-Peak Specialist occupies a lower position, at 0.114 and 0.242, respectively. Dinner-Peak Core retains the lowest stacked-state median, at 0.304 and 0.158. These comparisons locate agreement in relative order volume, bundling, and route length.

The temporal and spatial panels add a more differentiated picture. Lunch-Peak Specialist remains low in work-span and pickup-area rank, and Full-time Workhorse has similar median pickup-area ranks (0.568 and 0.575). Dinner-Peak Core shifts upward in work span and pickup area: its area median moves from the 0.213 ground-truth percentile to 0.625 in simulation. This rider-level difference coexists with its small pooled-area error in Table~\ref{tab:validation_results}, showing how aggregate reach and individual footprints capture different spatial structures. Percentiles summarize each source's own cohort, including its persona composition. Together, the untransformed table and percentile panels identify which behavioral dimensions retain their empirical ordering and which have different distributions under the 60-rider configuration.

\subsection{Fleet-Level Temporal Coordination}
\begin{figure*}[!t]
\centering
\includegraphics[width=0.98\textwidth]{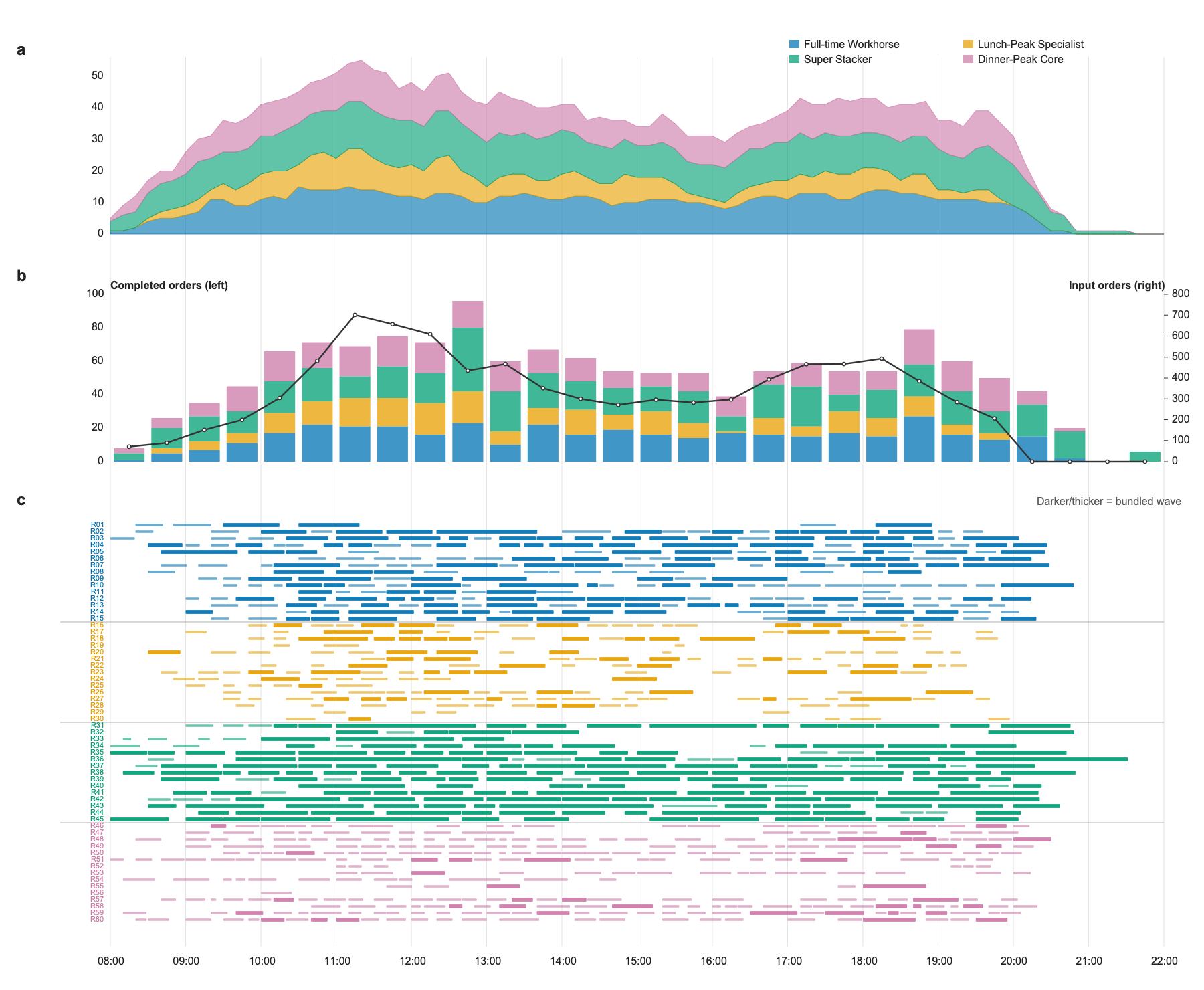}
\caption{Temporal organization of the 60-rider LLM-DR run. (a) Riders executing waves, observed every 10 minutes. (b) Completed orders per 30-minute interval (left axis) and input orders (black line, right axis). (c) Executed-wave timelines for all riders; darker, thicker segments indicate multi-order waves. Persona colors are shared across panels. Completion peaks later than input demand and service occupancy.}
\label{fig:temporal_outcomes}
\end{figure*}

Fig.~\ref{fig:temporal_outcomes} separates three temporal layers of the same 60-rider simulation. Panel (a) counts riders with an executing wave at 10-minute observation points, disaggregated by persona, and thus measures service concurrency. Panel (b) counts orders whose waves finish within each 30-minute interval and overlays all orders entering the frozen input stream in that interval on a separate right axis. Panel (c) shows every rider's wave start and finish times, ordered by persona, with darker and thicker segments denoting multi-order waves. The three panels respectively represent system occupancy, realized output relative to input pressure, and the individual action sequences from which the aggregate curves emerge.

The temporal layers peak at different times. Input demand reaches 701 orders between 11:00 and 11:30, while the number of simultaneously delivering riders peaks at 55 at 11:20. Completed output reaches its maximum later, at 96 orders between 12:30 and 13:00. This displacement reflects the simulated delivery pipeline: interval-batched arrivals enter the shared pool, accepted waves occupy riders for routed travel and service time, and all orders in a wave register completion at its end. Backlog, route duration, bundling, and the dispatch rule jointly shape the peak offset. Separating demand, concurrent labor, and realized output therefore reveals distinct temporal processes within the same action system.

Panel (c) exposes the microstructure behind that pipeline. Service is intermittent at the rider level even when fleet concurrency remains high, and single-order and bundled waves interleave rather than forming separate operating regimes. Work continues after the input window closes at 20:00, with residual completions extending toward 22:00; the simulation therefore preserves carryover work instead of truncating execution at the demand cutoff. Persona-colored lanes differ in the timing and density of their waves, but they also overlap through much of the day. As with the spatial results below, persona is expressed as a tendency within a shared environment rather than as a deterministic schedule.

\subsection{Activity Spaces and Spatiotemporal Organization}
Fig.~\ref{fig:temporal_outcomes} establishes when riders enter and leave routed service; the remaining figures examine where that service occurs. Activity space provides a bridge between human geography and agent-based delivery simulation. In time geography, mobility is not only a sequence of coordinates but the realized subset of urban opportunities that a person can reach under temporal, coupling, and physical constraints \cite{hagerstrandWhatPeopleRegional1970,millerMeasurementTheoryTime2005}. Activity-space measures summarize the locations and paths repeatedly incorporated into everyday behavior \cite{schonfelderActivitySpacesMeasures2003}. For platform delivery, these constraints are jointly social and computational: order release times and deadlines bound when movement is useful; accepted bundles couple a rider to several pickup and drop-off commitments; and the road network constrains how those commitments can be executed. We operationalize each simulated rider's activity space as the convex hull of all wave starts, pickups, and drop-offs. It is therefore a descriptive footprint of realized service opportunities, rather than a claim about every location the rider could have reached.

\begin{figure*}[!t]
\centering
\includegraphics[width=0.98\textwidth]{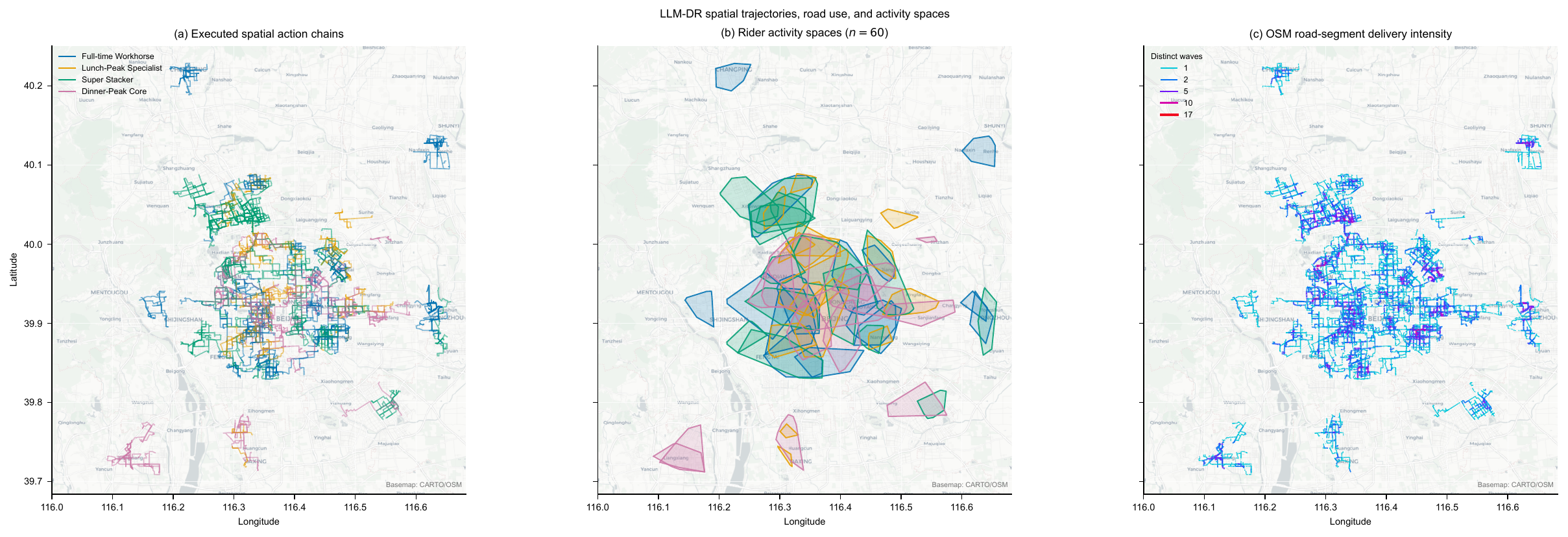}
\caption{Spatial organization of the 60-rider LLM-DR run on a common CARTO/OSM basemap and extent. (a) Executed road-network action chains, colored by persona. (b) Rider activity spaces, measured as convex hulls of wave starts, pickups, and drop-offs; darker overlaps indicate shared coverage. (c) Road-edge delivery intensity, counting distinct executed waves on each edge (blue to red). The personas share a central service core and common road corridors while showing different peripheral reach.}
\label{fig:spatial_outcomes}
\end{figure*}

Fig.~\ref{fig:spatial_outcomes} connects three spatial scales of the same simulation. Panel (a) shows the executed action chains after accepted decisions have been validated and routed. Panel (b) overlays the activity spaces of all 60 riders. Panel (c) aggregates the executed polylines to OSM road edges; intensity is the number of distinct executed delivery waves traversing each edge, not general traffic volume. The maps reveal a nested structure: activity spaces from all four personas intersect in central Beijing, where executed waves are also concentrated on a shared set of road corridors, whereas peripheral footprints are more separated and supported by locally distinct route segments. The core is thus characterized by shared spatial coverage and repeated infrastructure use, while the periphery is characterized by sparse, spatially discontinuous service territories.

This pattern is more informative than route density alone. The shared order stream creates a common spatial attraction field, persona-conditioned decisions determine when riders enter that field and whether they extend a wave through bundling, and the street network channels otherwise different choices onto common edges. Repetition converts these micro-level choices into a meso-level geography of overlapping activity spaces and concentrated corridor use. At the same time, the extensive cross-persona overlap in panel (b) cautions against interpreting persona labels as four geographically segregated rider populations. The present figure supports an emergent pattern of shared cores and differentiated spatial reach; a causal claim that persona alone produces those footprints would require matched demand exposure and repeated simulations.

\begin{figure*}[!t]
\centering
\includegraphics[width=0.92\textwidth]{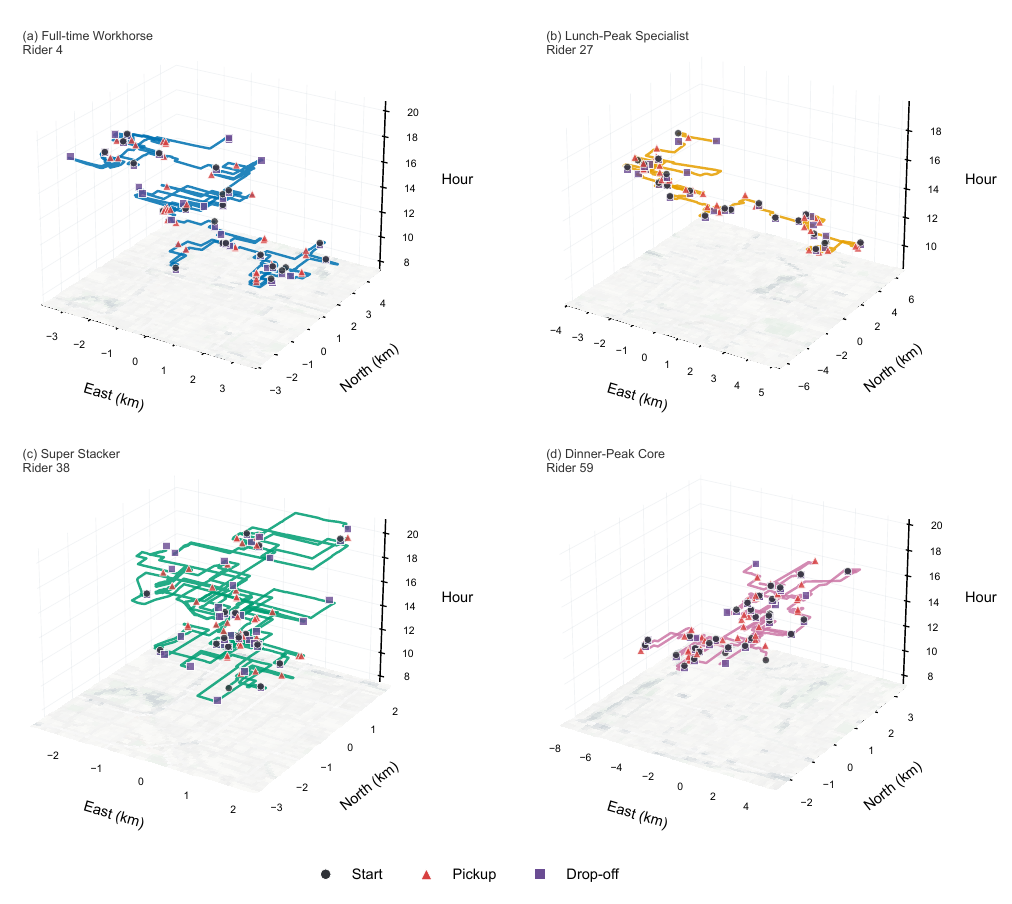}
\caption{Representative three-dimensional action chains from the 60-rider LLM-DR run, showing the most active rider in each persona. Lines trace route-tool-supported movement; markers identify start, pickup, and drop-off actions. Horizontal axes show local east and north distance (km), and the vertical axis shows time of day. The examples illustrate differences in service timing, multi-stop structure, and spatial reach.}
\label{fig:trajectory3d}
\end{figure*}

Fig.~\ref{fig:trajectory3d} adds time to this spatial organization for the most active rider in each persona group. Each polyline segment follows a validated decision through route-tool invocation, pickup, and drop-off, so vertical progression represents the accumulation of executed actions over the service day. The Full-time Workhorse example distributes actions across a broad temporal span and repeatedly returns to active delivery areas. The Lunch-Peak Specialist concentrates its action chain within a narrower interval. The Super Stacker forms longer multi-stop loops with dense pickup and drop-off sequences, corresponding to its high bundle size and stacked-wave ratio in Table~\ref{tab:main_results}. The Dinner-Peak Core shifts its action sequence later in the day and retains comparatively simple waves. Spatial overlap therefore does not imply identical spatiotemporal behavior: riders can use the same core at different times, remain there for different durations, and connect it to different peripheral areas. Together, Figs.~\ref{fig:spatial_outcomes} and \ref{fig:trajectory3d} show how temporal preference, bundling, and network-constrained execution jointly produce the fleet's realized activity system.

\section{Discussion}
\subsection{Heterogeneous Rider Responses}
The results support treating on-demand delivery as a coupled human-algorithm system. Under the same order stream and dispatch rule, the four persona groups produce different acceptance, bundling, and completion profiles (Table~\ref{tab:main_results}). This interaction between rider strategy and platform conditions is consistent with the machine-behavior view that algorithms and human actors should be studied jointly \cite{rahwanMachineBehaviour2019}. Persona grounding supplies an empirical prior for rider strategy, while the prompt, routing interface, and action log expose how an offer becomes an executed wave. Throughput alone would miss much of this variation. Comparing temporal, spatial, distance, and bundle distributions provides a separate check on whether operational outcomes retain empirical behavioral structure \cite{hofmanIntegratingExplanationPrediction2021}.

LLM-DR provides a rider-response model for studying dispatch systems. The present comparison isolates acceptance and bundling policies within one transparent assignment mechanism. Combining these agents with learned dispatch and matching methods would enable experiments on the interaction between platform objectives and heterogeneous rider responses \cite{chenImitationLearningEnhanced2022,chenOrderDispatchingGNN2024}.

Proposal-to-assignment differences locate rider interaction in a concrete platform mechanism: several riders can select an order, while only one receives it. Rider strategy therefore operates within an opportunity set shaped by others' previous assignments and current competing bids. This feedback links heterogeneous acceptance and bundling decisions to the work that is ultimately executed in the shared environment.

\subsection{Collective Spatiotemporal Organization}
The temporal and activity-space analyses extend this argument from individual actions to collective spatiotemporal organization. Fig.~\ref{fig:temporal_outcomes} distinguishes the demand clock, service-occupancy clock, and completion clock, while Fig.~\ref{fig:spatial_outcomes} connects realized service to individual footprints and edge-level infrastructure use. An LLM decision proposes an order set; identifier validation and conflict resolution produce an assignment; route execution occupies a rider in time and maps the action to street edges; and accumulated waves form overlapping activity spaces. Central overlap and displaced temporal peaks summarize the resulting interaction among shared demand, rider choices, service duration, and network constraints. The paired temporal and spatial outcomes expose how repeated individual actions accumulate into common corridor use and fleet-level rhythms.

\subsection{Scope and Limitations}
The experiment evaluates conditional single-day reconstruction with 60 agents equally distributed across four personas. The reference window contains 705 distinct couriers, whose daily participation differs from concurrent supply. Completion coverage is 16.5\%, so throughput, competition, and backlog reflect the chosen demand-to-supply configuration. Persona extraction and evaluation share data, and prompts contain statistics related to validation indicators. Availability-matched fleets, alternative persona compositions, held-out days, and component ablations would extend tests of environmental effects, temporal transfer, and agent capabilities.

Execution uses 10-minute order batches, idle-rider acceptance, and fixed pickup-then-drop-off waves. Orders become visible at their batch start, and the on-time proxy assigns each order the wave-end timestamp. The 120-second service allowance combines merchant waiting and customer handoff. These timing and assignment conventions shape workload and deadline outcomes. Event-time release, individual drop-off timestamps, mid-wave acceptance, route insertion, and joint capacity and deadline checks would support finer-grained modeling of rolling commitments.

Simulated earnings use an assumed distance-based reward schedule, while rider availability is represented within a single order pool. Recorded compensation and cross-platform assignments would support reward-sensitivity and multi-platform analyses. Convex-hull footprints are sensitive to peripheral actions; network-constrained or time-windowed measures could complement the road-edge and three-dimensional action-chain views.

\section{Conclusion}
This paper introduced LLM-DR, an empirical persona-grounded agentic LLM framework for simulating heterogeneous delivery rider behavior. The framework extracts rider personas, embeds them in structured LLM decision prompts, augments rider actions with a cycling-routing tool, and executes accepted waves in a dispatch simulation driven by real orders. In a controlled 60-rider reconstruction using 8,656 Beijing orders, the four persona groups generated distinct order-volume, reward, bundling, mileage, and wave-end deadline profiles. Super Stacker agents produced the highest order volume and strongest bundling, while Lunch-Peak Specialist and Dinner-Peak Core agents showed different temporal profiles. Fleet-level analysis distinguishes input demand, concurrent service, and wave completion, while activity spaces and road-edge intensity connect rider actions to collective urban structure. The single-day validation examines how data-derived personas carry through to executed actions and compares their order-time, route, and spatial distributions with empirical references. LLM-DR thus provides an executable setting for examining heterogeneous rider responses and their aggregate mobility outcomes under explicit fleet, dispatch, reward, and route-execution assumptions.

\IEEEtriggeratref{40}
\bibliographystyle{IEEEtran}
\bibliography{references}

\clearpage
\appendices
\section{Prompt Protocol}
To make the agent action protocol reproducible, this appendix reports the actual prompt template used by LLM-DR, in line with prompt-template reporting practices in LLM mobility-agent studies \cite{wangLargeLanguageModels2024}. Runtime variables generated by the simulator are replaced with angle-bracket placeholders. The final experiment used the route-tool version of the candidate-order block, where road-network distance and travel time are supplied by the bicycling routing tool before the LLM selects an action.

\smallskip
\noindent\fbox{%
\begin{minipage}{0.94\linewidth}
\scriptsize
\ttfamily
\raggedright
\textbf{Base instruction}\par
You are a delivery rider agent. Return concise strict JSON only.
\end{minipage}}

\smallskip
\noindent\fbox{%
\begin{minipage}{0.94\linewidth}
\scriptsize
\ttfamily
\raggedright
\textbf{Persona, state, and memory prompt}\par
\# ROLE AND GOAL\par
You are an AI agent acting as a delivery rider in a realistic simulation of Beijing's on-demand delivery market.\par
Make a human-like decision according to your rider persona and current operating context.\par
\smallskip
\# PERSONA PROFILE\par
Persona: \textless PERSONA\_NAME\textgreater\par
Description: \textless PERSONA\_DESCRIPTION\textgreater\par
Empirical statistics: \textless PERSONA\_STATS\_JSON\textgreater\par
\smallskip
\# CURRENT STATE\par
Current time: \textless YYYY-MM-DD HH:MM:SS TIMEZONE\textgreater\par
Current location: \textless RIDER\_LOCATION\_WKT\textgreater\par
Today's earnings so far: CNY \textless TOTAL\_EARNINGS\textgreater\par
Today's mileage so far: \textless TOTAL\_MILEAGE\_KM\textgreater km\par
Routing tool: estimates road-network travel distance and time between candidate waypoints.\par
\smallskip
\# MEMORY\par
\textless IF NO HISTORY: No completed orders yet.\textgreater\par
\textless ELSE: You have completed N orders, earned CNY TOTAL\_EARNINGS, and traveled TOTAL\_MILEAGE\_KM km.\textgreater\par
- Order \textless ORDER\_ID\textgreater: fee \textless FEE\textgreater, duration \textless DURATION\_S\textgreater s, \textless on time OR late\textgreater.\par
\textless repeat for up to five recent completed orders\textgreater
\end{minipage}}

\smallskip
\noindent\fbox{%
\begin{minipage}{0.94\linewidth}
\scriptsize
\ttfamily
\raggedright
\textbf{Candidate-order prompt}\par
\# AVAILABLE ORDERS\par
\textless for each candidate order i\textgreater\par
i. Order ID: \textless ORDER\_ID\textgreater\par
\ \ \ - Fee: CNY \textless FEE\textgreater\par
\ \ \ - Route-tool distance to pickup: \textless TO\_PICKUP\_KM\textgreater km\par
\ \ \ - Route-tool distance pickup-to-delivery: \textless DELIVERY\_KM\textgreater km\par
\ \ \ - Route-tool travel time: \textless TO\_PICKUP\_MIN\textgreater min to pickup; \textless DELIVERY\_MIN\textgreater min pickup-to-delivery\par
\ \ \ - Pickup: \textless PICKUP\_POINT\_WKT\textgreater\par
\ \ \ - Delivery: \textless DELIVERY\_POINT\_WKT\textgreater\par
\ \ \ - Promised delivery time: \textless HH:MM:SS\textgreater
\end{minipage}}

\smallskip
\noindent\fbox{%
\begin{minipage}{0.94\linewidth}
\scriptsize
\ttfamily
\raggedright
\textbf{Decision protocol and response schema}\par
\# DECISION WORKFLOW\par
Think through four concise steps:\par
1. Perception: summarize the offers.\par
2. Recall \& Evaluation: compare the offers with the persona.\par
3. Planning \& Bundling: decide whether a single order, a bundle, or rejection is best.\par
4. Decision: list the accepted order IDs.\par
\smallskip
\# OUTPUT FORMAT\par
Return only this JSON object:\par
\{\par
\ \ "accepted\_order\_ids": [integer order IDs chosen from AVAILABLE ORDERS, or []],\par
\ \ "reasoning\_steps": [\par
\ \ \ \ "Perception: ...",\par
\ \ \ \ "Recall \& Evaluation: ...",\par
\ \ \ \ "Planning \& Bundling: ...",\par
\ \ \ \ "Decision: ..."\par
\ \ ]\par
\}
\end{minipage}}

\smallskip
\noindent\fbox{%
\begin{minipage}{0.94\linewidth}
\scriptsize
\raggedright
\textbf{Placeholder instantiation.}
The following example instantiates every angle-bracket placeholder using the Full-time Workhorse case for rider 1 at 08:20 in the main simulation.
\par\smallskip
\texttt{\textless PERSONA\_NAME\textgreater} = Full-time Workhorse.\par
\texttt{\textless PERSONA\_DESCRIPTION\textgreater} = You are an experienced full-time delivery rider. You value stable daily earnings, steady order completion, and broad spatial coverage across the day.\par
\texttt{\textless PERSONA\_STATS\_JSON\textgreater} = a JSON-style dictionary such as \{num\_waves: 4.80; total\_orders: 15.62; work\_hr: 8.01; lunch\_ratio: 0.36; dinner\_ratio: 0.27; stacked\_order\_ratio: 0.26; avg\_wave\_distance\_m: 7046.50; activity\_area\_km2: 7.18; rider\_speed: 5.44; rider\_max\_load: 9.52\}.\par
\texttt{\textless YYYY-MM-DD HH:MM:SS TIMEZONE\textgreater} = 2020-02-15 08:20:00 +08:00.\par
\texttt{\textless RIDER\_LOCATION\_WKT\textgreater} = POINT (116.645782 39.904041).\par
\texttt{\textless TOTAL\_EARNINGS\textgreater} = 0.00 before the first accepted order; later memory example: 6.00.\par
\texttt{\textless TOTAL\_MILEAGE\_KM\textgreater} = 0.00 before the first accepted order; later memory example: 5.04.\par
\textless IF NO HISTORY: No completed orders yet.\textgreater{} = No completed orders yet.\par
\textless ELSE: You have completed N orders, earned CNY TOTAL\_EARNINGS, and traveled TOTAL\_MILEAGE\_KM km.\textgreater{} = You have completed 1 orders, earned CNY 6.00, and traveled 5.04 km.\par
\texttt{\textless ORDER\_ID\textgreater} = 6000000000000000215.\par
\texttt{\textless FEE\textgreater} = 6.0.\par
\texttt{\textless DURATION\_S\textgreater} = 1329.\par
\textless on time OR late\textgreater{} = on time.\par
\textless repeat for up to five recent completed orders\textgreater{} = - Order 6000000000000000215: fee 6.0, duration 1329s, on time.\par
\textless for each candidate order i\textgreater{} = one rendered candidate-order item, e.g., ``1. Order ID: 6000000000000000215''.\par
\texttt{\textless TO\_PICKUP\_KM\textgreater} = 1.98; \texttt{\textless DELIVERY\_KM\textgreater} = 3.06.\par
\texttt{\textless TO\_PICKUP\_MIN\textgreater} = 7.9; \texttt{\textless DELIVERY\_MIN\textgreater} = 12.2.\par
\texttt{\textless PICKUP\_POINT\_WKT\textgreater} = POINT (116.642803 39.920638).\par
\texttt{\textless DELIVERY\_POINT\_WKT\textgreater} = POINT (116.627457 39.919826).\par
\texttt{\textless HH:MM:SS\textgreater} = 10:45:00.
\end{minipage}}

\section{Representative Agent Action Outputs}
The following boxes report four representative action cases from the main LLM-DR simulation. Each case gives the context and reproduces the model's \texttt{reasoning\_steps}; LaTeX escaping and line breaks are applied, and order identifiers are pseudonymized in this release copy. These reported rationales support auditability of the agent action protocol as visible model outputs.

\begingroup
\setlength{\fboxsep}{2pt}

\smallskip
\noindent\fbox{%
\begin{minipage}{0.97\linewidth}
\tiny
\raggedright
\textbf{Full-time Workhorse, rider 1, 08:20.}
\par\smallskip
\textbf{Case context:} available orders: [6000000000000000215]; selected orders: [6000000000000000215].
\par\smallskip
\textbf{Agent output:}
\par\smallskip
\textbf{Perception:} Only one order available. Fee: CNY 6.0. Distance to pickup: 1.98 km, pickup-to-delivery: 3.06 km. Total distance: $\sim$5.04 km. Travel time: 7.9 min to pickup + 12.2 min delivery = $\sim$20.1 min. Promised delivery time: 10:45, current time 08:20, so plenty of time.
\par\smallskip
\textbf{Recall \& Evaluation:} As a Full-time Workhorse, I value steady earnings and broad spatial coverage. The fee is moderate, and the total distance is within my typical wave distance ($\sim$7 km). The time commitment is low, leaving room for more orders later. Accepting this order aligns with my goal of stable daily earnings.
\par\smallskip
\textbf{Planning \& Bundling:} Only one order available, so no bundling possible. Accepting it is straightforward.
\par\smallskip
\textbf{Decision:} Accept the single order to start the day's earnings.
\end{minipage}}

\smallskip
\noindent\fbox{%
\begin{minipage}{0.97\linewidth}
\tiny
\raggedright
\textbf{Lunch-Peak Specialist, rider 16, 09:50.}
\par\smallskip
\textbf{Case context:} available orders: [6000000000000000651, 6000000000000000470]; selected orders: [6000000000000000470].
\par\smallskip
\textbf{Agent output:}
\par\smallskip
\textbf{Perception:} Two orders available. Order 1: CNY 6.0, pickup 1.96 km away, delivery 5.35 km, total distance $\sim$7.31 km, travel time $\sim$29.2 min, promised by 10:56. Order 2: CNY 6.0, pickup 0.54 km away, delivery 2.64 km, total distance $\sim$3.18 km, travel time $\sim$12.8 min, promised by 11:15.
\par\smallskip
\textbf{Recall \& Evaluation:} As a Lunch-Peak Specialist, I prefer short, dense, efficient orders during lunch rush. It's 09:50, still early for peak lunch, but I can take a quick nearby order. Order 2 is very close (0.54 km pickup) and short total distance (3.18 km), fitting my preference for nearby efficient trips. Order 1 is longer (7.31 km total) and less efficient. My persona avoids work that doesn't fit peak window; Order 2 is quick and won't tie me up long before lunch peak.
\par\smallskip
\textbf{Planning \& Bundling:} Bundling both would increase total distance and time, risking missing better lunch orders. Order 2 alone is optimal---quick, nearby, and leaves me free for lunch peak. Order 1 is rejected due to distance and lower efficiency.
\par\smallskip
\textbf{Decision:} Accept only Order 2.
\end{minipage}}

\smallskip
\noindent\fbox{%
\begin{minipage}{0.97\linewidth}
\tiny
\raggedright
\textbf{Super Stacker, rider 31, 10:10.}
\par\smallskip
\textbf{Case context:} available orders: [6000000000000000823, 6000000000000001092, 6000000000000000536, 6000000000000000814]; selected orders: [6000000000000000823, 6000000000000001092].
\par\smallskip
\textbf{Agent output:}
\par\smallskip
\textbf{Perception:} Four orders available. Order 1: CNY 3.0, short distance (0.90 km to pickup, 1.23 km delivery), early deadline 10:40. Order 2: CNY 6.0, moderate distance (1.19 km to pickup, 1.68 km delivery), deadline 10:52. Order 3: CNY 6.0, longer distance (1.64 km to pickup, 2.88 km delivery), deadline 11:12. Order 4: CNY 6.0, similar pickup to Order 3 but longer delivery (3.43 km), deadline 11:15.
\par\smallskip
\textbf{Recall \& Evaluation:} As Super Stacker, I prefer multi-order bundles and spatially coherent routes. Order 1 and Order 2 have nearby pickups (POINT (116.336902 40.032579) and POINT (116.337449 40.032518)) and deliveries in similar areas (Order 1 delivery: POINT (116.330267 40.034286), Order 2 delivery: POINT (116.323907 40.035646)). They can be stacked efficiently. Order 3 and Order 4 share the same pickup location but are farther away and have longer delivery distances; they could be bundled together but would take more time and distance, reducing efficiency. Order 1 fee is low but can be combined with Order 2 for a total of CNY 9.0 with minimal extra distance.
\par\smallskip
\textbf{Planning \& Bundling:} Bundle Order 1 and Order 2. Pickup Order 1 first (closer), then Order 2 (very close), then deliver Order 1, then Order 2. Total distance: to pickup Order 1 0.90 km, to pickup Order 2 $\sim$0.05 km (since pickups are adjacent), to delivery Order 1 1.23 km, to delivery Order 2 $\sim$0.8 km (estimated from Order 1 delivery to Order 2 delivery). Total $\sim$2.98 km, earning CNY 9.0. Both deadlines are feasible: Order 1 by 10:40 (30 min from now), Order 2 by 10:52 (42 min). Reject Orders 3 and 4 as they would increase distance significantly and risk delays.
\par\smallskip
\textbf{Decision:} Accept Order 1 and Order 2 as a bundle.
\end{minipage}}

\smallskip
\noindent\fbox{%
\begin{minipage}{0.97\linewidth}
\tiny
\raggedright
\textbf{Dinner-Peak Core, rider 46, 08:30.}
\par\smallskip
\textbf{Case context:} available orders: [6000000000000000290, 6000000000000000558]; selected orders: [6000000000000000290].
\par\smallskip
\textbf{Agent output:}
\par\smallskip
\textbf{Perception:} Two orders available. Order 1: CNY 3.0, 2.71 km to pickup, 1.19 km delivery, total $\sim$15.6 min, promised by 09:04. Order 2: CNY 6.0, same pickup, 2.59 km delivery, total $\sim$21.1 min, promised by 09:08. Both pickups at same location.
\par\smallskip
\textbf{Recall \& Evaluation:} As a Dinner-Peak Core rider, I prefer direct, manageable orders and avoid overly complex bundles. It's 08:30, not my peak dinner time, but I can take a short order. Order 1 is short and simple, fits my style. Order 2 is longer but pays more. Bundling both would increase complexity and distance, which I typically avoid.
\par\smallskip
\textbf{Planning \& Bundling:} I could take both since pickup is same, but total distance becomes 2.71 + 1.19 + 2.59 = 6.49 km, which is above my average wave distance of $\sim$9 km but still manageable. However, my persona avoids overly complex bundles. Taking only Order 1 is simpler and aligns with my preference for direct orders. Order 2 alone is less attractive due to longer delivery distance. Rejecting both is not ideal since I have no earnings yet.
\par\smallskip
\textbf{Decision:} Accept only Order 1 as it's a short, direct order that fits my persona. Reject Order 2 to avoid complexity.
\end{minipage}}

\endgroup

\end{document}